\documentclass[english,twocolumn,twocolumn,showpacs,preprintnumbers,amsmath,amssymb,superscriptaddress]{revtex4}
\usepackage[T1]{fontenc}
\usepackage[latin1]{inputenc}
\usepackage{graphicx}

\makeatletter

\providecommand{\tabularnewline}{\\}

\makeatletter

\makeatletter

\usepackage{color}

\usepackage{dcolumn}

\usepackage{bm}

\newcommand{\pt}{ p_{\rm t}}
\newcommand{\ie}{{\it i.e.}}

\def\lsim{\mathrel{\rlap{\lower4pt\hbox{\hskip1pt$\sim$}}
    \raise1pt\hbox{$<$}}}         
\def\gsim{\mathrel{\rlap{\lower4pt\hbox{\hskip1pt$\sim$}}
    \raise1pt\hbox{$>$}}}         

\usepackage{babel}
\makeatother
\begin{document}

\title{Elliptic flow of thermal photons in Au+Au collisions at $\sqrt{s_{NN}}=200$
GeV}

\author{Fu-Ming Liu}

\affiliation{Institute of Particle Physics, Central China Normal University, Wuhan,
China}

\author{Tetsufumi Hirano}

\affiliation{Department of Physics, The University of Tokyo, 113-0033, Japan}

\author{Klaus Werner}

\affiliation{Laboratoire SUBATECH, University of Nantes - IN2P3/CNRS - Ecole desMines,
Nantes, France}

\author{Yan Zhu}

\affiliation{Institute of Particle Physics, Central China Normal University, Wuhan,
China}

\date{\today}

\begin{abstract}
The transverse momentum ($\pt$) dependence, the centrality dependence
and the rapidity dependence of the elliptic flow of thermal photons
in Au+Au collisions at $\sqrt{s_{NN}}=200$ GeV are predicted, based
on a three-dimensional ideal hydrodynamic description of the hot and
dense matter. The elliptic flow parameter $v_{2}$, i.e.~the second
Fourier coefficient of azimuthal distribution, of thermal photons,
first increases with $\pt$ and then decreases for $\pt>$ 2 GeV/$c$,
due to the weak transverse flow at the early stage. The $\pt$-integrated
$v_{2}$ first increases with centrality, reaches a maximum at about
50\% centrality, and decreases. The rapidity dependence of the elliptic
flow $v_{2}(y)$ of direct photons (mainly thermal photons) is very
sensitive to the initial energy density distribution along longitudinal
direction, which provides a useful tool to extract the realistic initial
condition from measurements.
\end{abstract}
\maketitle

\section{Introduction}

The deconfined and novel nuclear matter, the quark gluon plasma (QGP),
has been expected to appear in relativistic heavy ion collisions.
Various signatures have been proposed to verify its existence \cite{sign}.
Many experiments have been done so far to explore properties of the
QGP at the Relativistic Heavy Ion Collider (RHIC) and will be done
at the Large Hadron Collider (LHC). Experimental results indicate
that the QGP is created at RHIC \cite{rhic4}.

Collective flow, in particular elliptic flow, is an effective probe
to investigate bulk properties of the QGP in nucleus-nucleus collisions
at relativistic energies. In non-central collisions the overlapping
reaction zone of two colliding nuclei has an anisotropic shape, like
an almond, in the transverse plane. This leads to a preferred flow
direction parallel to the impact parameter, and therefore the initial
spatial anisotropy is carried over to momentum-space anisotropy \cite{v2}.
The QGP created at the early stage expands, cools down, and finally
forms a hadronic gas (HG) phase. Then hadrons still have strong rescatterings
until freeze-out. So those hadrons do carry information on the asymmetric
flow, but only concerning the freeze out surface.

Fortunately, in high energy heavy ion collisions, thermal photons
can be produced during the whole history of the evolution of the hot
and dense matter. Moreover the mean free path of photons is much larger
than the transverse size of the bulk matter. So thermal photons produced
in the interior of the plasma pass through the surrounding matter
without any interaction. As a result, thermal photons provide information
on flow asymmetries even from inside the bulk volume, not just its
surface.

In this paper, we investigate the relation between the observables
of thermal photons and the space-time evolution of the hot and dense
matter for exploring QGP properties, based on a fully three dimensional
(3D) ideal hydrodynamic description of expansion of the matter. Toward
establishment of the relation, we calculate the second Fourier coefficient
of azimuthal momentum distribution, the so-called elliptic flow coefficient
$v_{2}$, for thermal photons in Au+Au collisions at $\sqrt{s_{NN}}=200$
GeV at various centralities (0-70\%).

In Sec.~2 we will briefly review the space-time evolution of the
hot and dense matter using the 3D ideal hydrodynamics and the basic
formula for the production of thermal photons. In Sec.~3, we will
show our results on $v_{2}$ of thermal photons in Au+Au collisions
at $\sqrt{s_{NN}}=200$ GeV, its transverse momentum, centrality
dependences and rapidity dependence. Section 4 is devoted to discussion and summary of our
results.

\section{Bulk evolution, thermal photons, and elliptic flow}

\begin{figure}
\begin{centering}
\includegraphics[scale=0.8]{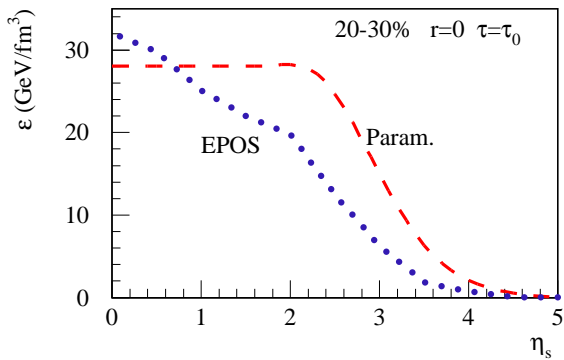} 
\par\end{centering}
\caption{\label{yeps}(Color online) energy density as a function of space-time
$\eta_s$;Dashed lines from parameterized initial condition and dotted
lines from EPOS initial condition.}
\end{figure}

In our calculation, a full 3D ideal hydrodynamic calculation \cite{hyd1,hyd2}
is employed to describe the space-time evolution of the hot and dense
matter created in Au+Au collisions at RHIC. 
The local thermal equilibrium is assumed to 
be reached at the initial time $\tau_{0}=0.6$ fm/$c$.
The transverse flow is assumed to be zero 
at $\tau_{0}$.
We consider two scenarios to obtain entropy density (or energy density) at $\tau_{0}$:
one is a parameterization based on Glauber model\cite{hyd1,hyd2}, 
and the other is based on flux tubes (string picture) realized 
in the EPOS model\cite{klaus}.
For these two initial conditions, 
 we take the same hydrodynamic equations and equations of state. The
initial energy density from the two scenarios has been plotted as
a function space-time rapidity $\eta_s$ 
in Fig.\ref{yeps}, where the dashed
lines refer to the parameterized initial conditions, and dotted lines
to the  EPOS case. One can see that the energy density decreases
very rapidly with $\eta_s$ in EPOS, while it has a
plateau for $|\eta_s|<2.5$ using parameterized initial conditions. In two
dimensional hydrodynamics this plateau is assumed for the whole $\eta_s$ region.
The EPOS initial condition gives many similar results compared to
the parameterized ones, $\ie,$ the measured
$\pt$-spectra of direct photons can also be well reproduced, and
the same $\pt$-dependence and centrality dependence of elliptic flow
of thermal photons has been observed. But there are also striking differences, to be
discussed later.

For both scenarios,
the impact parameters corresponding to different centralities in Au+Au
collisions at RHIC are 3.2,
5.5, 7.2, 8.5, 9.7, 10.8, and 11.7 fm for 0-10\%, 10-20\%, $\cdots$\,
and 60-70\% centrality, respectively. 
The evolution of the fluid is
governed by conservation laws of energy and momentum \begin{eqnarray}
\partial_{\mu}T^{\mu\nu}=0\end{eqnarray}
 with the energy-momentum tensor \begin{eqnarray}
T^{\mu\nu}=(e+P)u^{\mu}u^{\nu}-Pg^{\mu\nu}\end{eqnarray}
for a perfect fluid. Here, $e$, $P$, and $u^{\mu}$ are energy density,
pressure, and local four fluid velocity, respectively, which are related 
through the equation of state (EOS).
The above equations are solved in full 3D space ($\tau$, $x$, $y$, $\eta_{s}$) where
$\tau$, $\eta_{s}$, $x$, and $y$ are the proper time, space-time
rapidity, the two transverse coordinates along the impact parameter
and perpendicular to the reaction plane, respectively. Since the net-baryon
density is small near the mid-rapidity at RHIC, it is neglected. We
also neglect the dissipative effects in the space-time evolution.
The critical temperature of a first order phase transition between
the QGP phase and the hadron phase is fixed at $T_{c}=170$ MeV. The
 EOS for the QGP phase is \begin{eqnarray}
P=\frac{1}{3}(e-4B),\end{eqnarray}
 where the bag constant is tuned to $B^{\frac{1}{4}}=247.19$ MeV
so that pressure of the QGP phase with three flavors is matched to
that of HG phase (up to mass of $\Delta(1232)$) at $T_{c}$. In the
hadronic phase, we employ two resonance gas models: partial chemical
equilibrium (PCE) and full chemical equilibrium (FCE) \cite{hyd2}.
In the PCE model, the particle ratios including contribution from
resonance decays are assumed to keep during the hadronic expansion,
which is required from the data of particle ratios. The PCE model
is our standard choice to calculate the photon spectra in the following.
For the purpose of comparison, we also employ the conventional resonance
gas model in which chemical equilibrium is maintained during the hadronic
expansion. In both models, the hadronic matter is kinetically and
thermally frozen at $e^{\mathrm{th}}=0.08$ GeV/fm$^{3}$. Corresponding
freezeout temperature is $T^{\mathrm{th}}\sim100$ (130) MeV for the
PCE (FCE) model, which is consistent with the $\pt$ slope of protons.

Transverse momentum spectra of thermal photons can be written as \begin{equation}
\frac{dN}{dy\, d^{2}p_{t}}=\int d^{4}x\,\Gamma(E^{*},T)\label{eq1}\end{equation}
 with $\Gamma(E^{*},T)$ being the Lorentz invariant thermal photons
emission rate which covers the contributions from the QGP phase \cite{resul}
and HG phase \cite{MYM}, $d^{4}x=\tau\, d\tau\, dx\, dy\, d\eta_{s}$
being the volume-element, and $E^{*}=p^{\mu}u_{\mu}$ the photon energy
in the local rest frame. Here $p^{\mu}$ is the photon's four momentum
in the laboratory frame, $T$ and $u^{\mu}$ are the temperature and
the local fluid velocity, respectively, to be taken from the hydrodynamic
calculations mentioned above. We only consider thermal photon radiation
from the region with energy density above $e^{\mathrm{th}}$.
For more details one can check \cite{Liu:2008eh} where the measured $p_{t}$ spectra 
of direct photons at different centralities in Au+Au collisions
are nicely reproduced.
   
The triple differential spectrum can be written as a Fourier series,
\begin{eqnarray}
\frac{d^{3}N}{dy\, d^{2}p_{t}}=\frac{d^{2}N}{2\pi p_{t}\, dp_{t}\, dy}\left(1+\sum_{n=1}^{\infty}2v_{n}\cos(n\phi)\right),\label{eqdn}\end{eqnarray}
 where $\phi$ is the azimuthal angle of photon's momentum with respect
to the reaction plane, which is defined to be the plane containing
the impact parameter and beam axis. The elliptic flow is quantified
by the second harmonic coefficient $v_{2}$ \begin{eqnarray}
v_{2}(\pt,y)=\frac{\int d\phi\cos(2\phi)d^{3}N/dy\, d^{2}p_{t}}{\int d\phi d^{3}N/dy\, d^{2}p_{t}}.\end{eqnarray}

The $p_{t}$ dependence of the triple differential spectra is strongly
affected by the flow $u$ through the argument $E^{*}=p^{\mu}u_{\mu}$
in the photon emission rate. 
In the local rest frame thermal photons of any given energy are emitted
isotropically. The isotropic distribution is Lorentz-boosted by flow
velocity $\vec{v}_{r}$. The azimuthal asymmetry of the transverse
components of the flow obviously ends up with an anisotropic momentum
distribution, which gives a finite $v_{2}$. Therefore both strength
and the anisotropy of transverse flow velocity are important to generate
the elliptic flow of thermal photons. To understand the thermal photon
$v_{2}$ quantitatively, we define the mean radial flow $\left\langle v_{r}\right\rangle $
and the mean anisotropy of flow $\left\langle v_{2}^{\mathrm{hydro}}\right\rangle $
as \begin{eqnarray}
\left\langle v_{r}\right\rangle  & = & \left\langle \sqrt{v_{x}^{2}+v_{y}^{2}}\right\rangle \label{eq:aveflow1}\\
\left\langle v_{2}^{\mathrm{hydro}}\right\rangle  & = & \left\langle \cos2\phi_{v}\right\rangle =\left\langle \frac{v_{x}^{2}-v_{y}^{2}}{v_{x}^{2}+v_{y}^{2}}\right\rangle ,\label{eq:aveflow2}\end{eqnarray}
 where $\left\langle \cdots\right\rangle $ stands for energy-density-weighted
space-time average, $v_{x}$ and $v_{y}$ are the flow velocity components
along $x$-axis and $y$-axis, respectively. We only consider the
region above $e=e^{\mathrm{th}}=0.08$ GeV/fm$^{3}$, where thermal
photons are produced in this model. %
\begin{table}[h]

\caption{\label{cent}The strength and the anisotropy of transverse flow at
each centrality.}

\begin{ruledtabular} \begin{tabular}{cccccccc}
Centrality(\%) &
0-10 &
10-20 &
20-30 &
30-40 &
40-50 &
50-60 &
60-70\tabularnewline
$\left\langle v_{r}\right\rangle $ &
0.114 &
0.122 &
0.123 &
0.117 &
0.109 &
0.0959 &
0.0804\tabularnewline
$\left\langle v_{2}^{\mathrm{hydro}}\right\rangle $ &
0.0417 &
0.103 &
0.154 &
0.188 &
0.212 &
0.222 &
0.240 \tabularnewline
\end{tabular}\end{ruledtabular} 
\end{table}

\begin{figure*}
\begin{centering}
\includegraphics[scale=0.8]{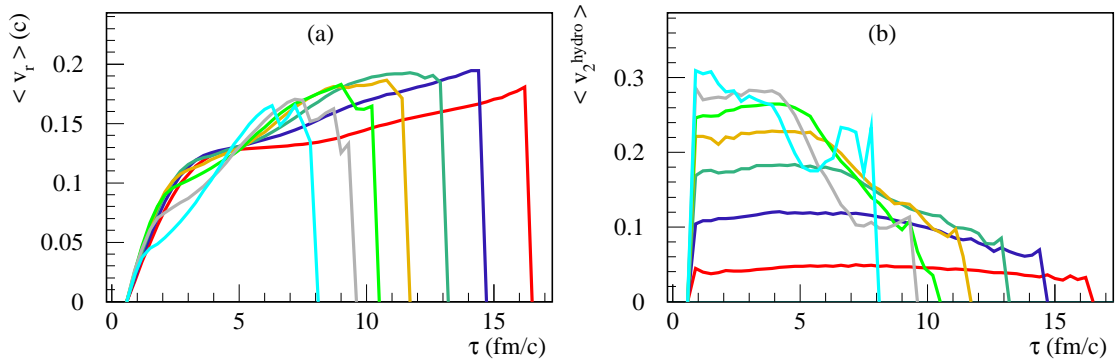} 
\par\end{centering}

\caption{\label{v2H} (Color online) Average radial flow velocity (a) and
average flow anisotropy (b) as a function of proper time. Different
curves correspond to 0-10\%, 10-20\%, ... 60-70\%, respectively. A
line with longer life time represents the corresponding result for
a more central collision. }
\end{figure*}

From Table~\ref{cent}, we see that the mean radial flow increases
with centrality from 0-10\% ($b=3.2$ fm) to 20-30\% ($b=7.2$ fm),
then decreases from 20-30\% ($b=7.2$ fm) to 60-70\% ($b=11.7$ fm).
On the other hand, the mean flow anisotropy increases with centrality
from 0-10\% to 60-70\% monotonically. 
One can easily understand the tendency when we investigate the time 
evolution of space-averaged results in Figs.~\ref{v2H} (a) and (b).
One should notice that, for all centralities, the average
radial flow is quite small before $\tau=1$ fm/$c$ during which most
of thermal photons in $4<\pt<6$ GeV/$c$ are produced.

\section{Results}

\begin{figure}
\includegraphics[scale=0.8]{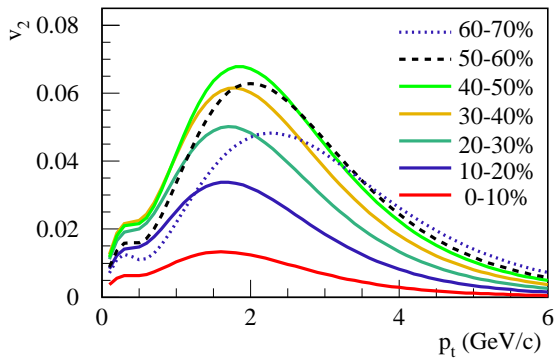}

\caption{\label{v2} (Color online) $v_{2}$ for thermal photons in Au+Au
collisions at $\sqrt{s_{NN}}=200$GeV is shown for various centralities
from 0 to 70\% in $0<\pt<6$ GeV/$c$. The solid lines from bottom
to top correspond to centralities 0-10\%, 10-20\%, 20-30\%, 30-40\%,
and 40-50\%, respectively. The dashed lines from top to bottom at
$\pt=$ 2 GeV/$c$ correspond to the centralities 50-60\% and 60-70\%. }
\end{figure}

In Fig.~\ref{v2}, the $v_{2}$ of thermal photons in $0<\pt<6$
GeV/$c$ is shown for various centralities from 0 to 70\% in Au+Au
collisions at $\sqrt{s_{NN}}=200$ GeV. The solid lines from bottom
to top refer to centralities 0-10\%, 10-20\%, 20-30\%, 30-40\%, and
40-50\%, respectively. The dashed lines from top to bottom at $\pt=$
2 GeV/$c$ refer to the centralities 50-60\% and 60-70\%.

For each centrality, the thermal $v_{2}$ increases then decreases
with increasing $p_{t}$ and a peak appears at $p_{t}\sim2$~GeV/$c$.
The same $p_{t}$ dependence has been predicted in the study with
2+1D ideal hydrodynamics \cite{ther}. The decrease of thermal photons'
$v_{2}$ at high $p_{t}$ can be explained as the weak transverse
flow at the early stage. Because at the higher $\pt$, the more fraction
of thermal photons are emitted from the hot matter at an early stage
for all centralities. $\ie$, at $\pt=3$ GeV/$c$, about 50\% thermal
photons are produced during $\tau\in(0.6,0.9)$fm/$c$ and at $\pt=4$
GeV/$c$, the fraction is about 70\%. In our 3+1D hydrodynamics, the
evolution of the radial flow is plotted in Fig.~\ref{v2H}(a). At
an early time near $\tau_{0}$, $\ie,$$\tau\in(0.6,0.9)$fm/$c$,
the transverse flow vanishes, and so does the elliptic flow of thermal
photons. Therefore, the decrease of thermal photons' $v_{2}(p_{t})$
at high $\pt$ reflects the weak transverse radial flow in the early
stage. Note that a similar non-trivial behavior is also seen in $\pt$
slope parameters of dimuon spectra as a function of invariant mass
\cite{alam}.

Similar to the 2+1D hydrodynamics\cite{ther}, small bumps appear
in the elliptic flow of thermal photons at $\pt$ close to zero. This
is because of the thermal photons are emitted from two phases, QGP
phase and hadronic phase at different time. In fact, 
thermal photon emission rate~\cite{resul} is not reliable at $\pt \rightarrow 0$.

\begin{figure}
\begin{centering}
\includegraphics[scale=0.8]{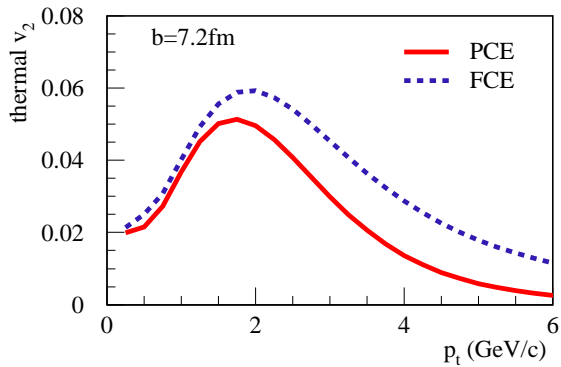} 
\par\end{centering}

\caption{\label{b72} (Color online) The elliptic flow of thermal photons
at $b=7.2$ fm. Solid (Dashed) line is the result from a partial chemical
equilibrium model (a full chemical equilibrium model). }
\end{figure}

One can also see an EOS dependence of thermal photon elliptic flow.
As an example, we change from the PCE model to the FCE model, which
has been employed in the conventional hydrodynamic calculations including
the thermal photon calculations in the earlier study \cite{ther}.
The change of the EOS is almost invisible in the $\pt$ spectra of
thermal photons due to the dominant contribution from the QGP phase.
However, $v_{2}(\pt)$ from the hadronic phase is about ten times
bigger than the one from the QGP phase. So a slight change in the
hadronic phase can be magnified and become quite visible in the elliptic
flow of thermal photons, as shown in Fig.~\ref{b72}. One can also
conclude that the elliptic flow of thermal photons is sensitive to
the EOS, particularly in the hadron phase. However, the effect of
different EOS we obtained here is quite similar to the effect from
the different formation time of quark gluon plasma \cite{Chatterjee:2008tp}.

\begin{figure*}
\begin{centering}
\includegraphics[scale=0.8]{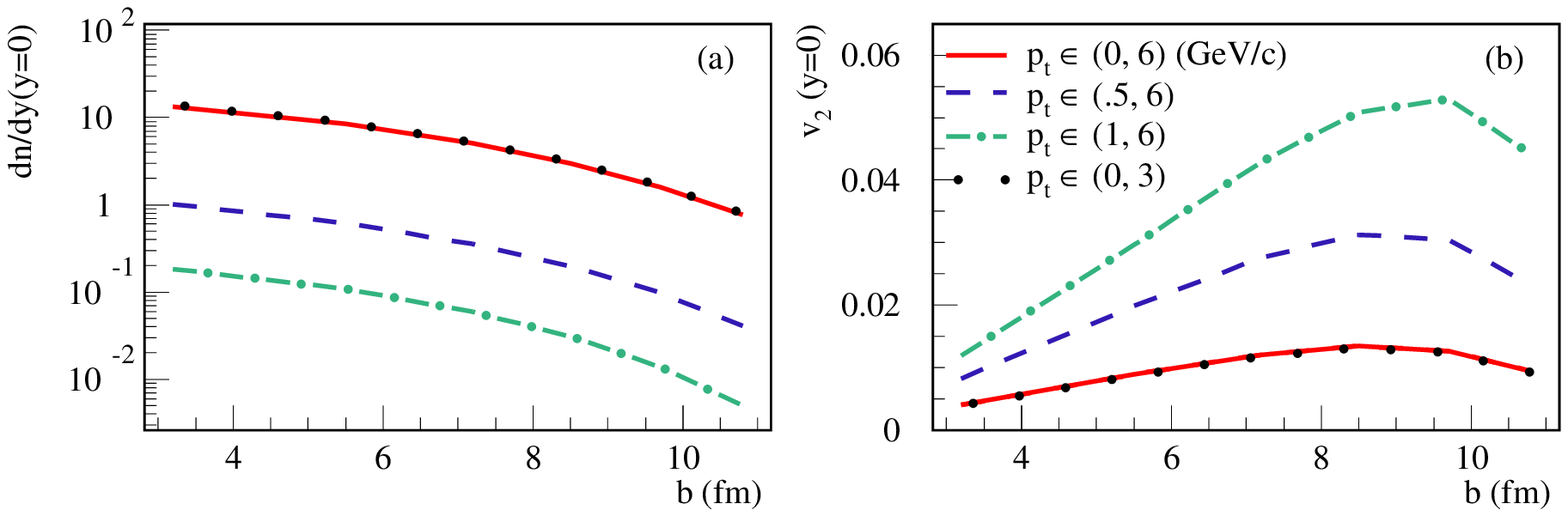} 
\par\end{centering}

\caption{\label{ptb}(Color online) (a) the yield and (b) the 
$v_{2}$ of thermal photons at midrapidity. Different curve types
represent different $\pt$-integral ranges, i.e. (0,3)GeV/$c$: dotted
lines; (0,6): solid lines; (0.5, 6): dashed lines; (1,6): dashed dotted
lines. }
\end{figure*}

The centrality dependence has been investigated
in Fig.\ref{ptb}, where (a) represents the yield and (b) the 
$v_{2}$ of thermal photons at midrapidity, and  where different types of
curves present different $\pt$ range. It is clear that
the yield of thermal photons decreases with centrality monotonously.
However, the elliptic flow of thermal photons does not change with
centrality monotonously. It reaches a maximum at impact parameter
b=9.7fm or 40-50\% centrality, then decreases for more central or
more peripheral collisions, due to the interplay between the strength
and the anisotropy of transverse flow. As we discussed in Sec.~2,
both the strength and the anisotropy of transverse flow are important
factors to generate the elliptic flow of thermal photons. The anisotropy
increases with decreasing centrality, but the strength of transverse
flow decreases, as shown in Table~\ref{cent}. 

In Fig.\ref{ptb} the $\pt$ ranges used are (0,3)GeV/$c$:
dotted lines; (0,6): solid lines; (0.5, 6): dashed lines; (1,6): dashed
dotted lines. One can see the yield and the $\pt$-integrate $v_{2}$
of thermal photons are not sensitive to the upper limit, c.f. the
integral range (0,3) and (0,6) GeV. This is clear because the $\pt$-spectrum
of thermal photons decrease almost exponentially and high $\pt$ region
contributes very little to the total yield. The $\pt$-spectrum of
direct photons is dominant by the thermal photons at low $\pt$region
and also decreases rapidly with $\pt$. So we can expect that the
$\pt$-integrated quantities of direct photons behaviors similarly
to those of thermal photons. On the other side, one can see that the
yield and the $\pt$-integrate $v_{2}$ of thermal photons are very
sensitive to the lower limit, due to the rapidly decreasing $\pt$-spectrum.
One should be careful when a comparison between the prediction and
the measurement is performed since the lower limit is experimentally
determined by the detectors.

\begin{figure*}
\begin{centering}
\includegraphics[scale=0.8]{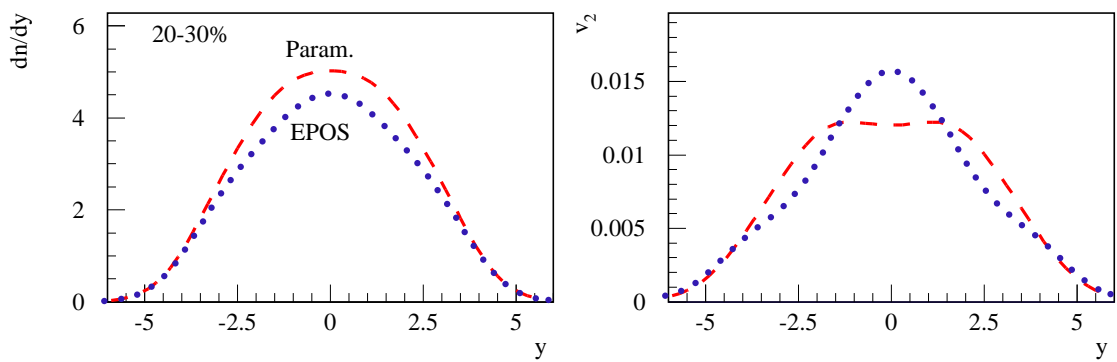} 
\par\end{centering}

\caption{\label{ydis}(Color online) (a): rapidity dependence of thermal photon
production. (b): rapidity dependence of the $\pt$-integrate $v_{2}$
of thermal photons. Dashed lines from parameterized initial condition
and dotted lines from EPOS. }
\end{figure*}

The above results are all at midrapidity. In the following we discuss the
rapidity dependence, for the two scenarios: initial
conditions based on EPOS flux tubes and the parameteraized one.
As already said, both scenarios give many similar results.
However, the rapidity
dependence of elliptic flow is very different:
in Fig.\ref{ydis}, the rapidity distributions of $dn/dy$ and the
$v_{2}$ of thermal photons produced from the two
kinds of initial conditions are shown. Dashed lines for parameterized
one and dotted lines for EPOS. One can see thermal photon yield $dn/dy$
has a similar dependence on rapidity $y$, but two very different
dependences of the elliptic flow $v_{2}(y)$ have been obtained. One
can see that  $v_{2}(y)$ shows almost the same shape as the 
initial $\epsilon(\eta_s)$: A rapid decrease of elliptic flow along
longitudinal direction is obtained from EPOS initial conditions, where
the energy density decreases very rapidly with $\eta_s$, while a plateau
of $v_{2}(y)$ at midrapidity region is obtained from the parameterized
initial conditions. A similar observation has already been made concerning
 the rapidity dependence of
elliptical flow of hadrons\cite{klaus}.

\section{Discussion and Conclusion}

Based on a fully three-dimensional ideal hydrodynamic description
of the evolution of hot and dense matter created in Au+Au collisions
at $\sqrt{s_{NN}}=200$ GeV, we found that the elliptic flow of thermal
photons depends non-trivially on not only the anisotropy but also
the strength of transverse flow.

At midrapidity, the $v_{2}$ of thermal photons increases with $\pt$
up to $p_{t}\sim2$ GeV/$c$, then decreases at higher $p_{t}$, similar
to the prediction from two-dimensional ideal hydrodynamics \cite{ther}.
The decrease of thermal photon $v_{2}$ at high $p_{t}$ reflects
the fact that at very early times, the transverse radial flow is weak.
So thermal photons can serve as a penetrating and unique probe for
the whole history of the evolution.

The $\pt$-integrated $v_{2}$ of thermal
photons increases with increasing impact parameter $b$ up to $\sim10$
fm (40-50\% centrality) and then decreases above. Thermal photons
from two different phases have the same centrality dependence. The
decrease of thermal photon $v_{2}$ at peripheral collisions is due
to the interplay between the strength and anisotropy of transverse
radial flow of the hot matter: As increasing impact parameter $b$,
the anisotropy increases, but the strength of transverse radial flow
decreases.

The rapidity dependence of elliptic flow of thermal photons $v_{2}(y)$
can ``remember'' the initial energy density along the longitudinal
direction $\epsilon_{0}(\eta_s)$, $\ie,$ a plateau at midrapidity
can be obtained from the parameterized initial condition based on Glauber model,
 while a rapid decrease of elliptic flow along rapidity
is obtained from EPOS initial conditions.

\begin{acknowledgments}
This work is supported by the Natural Science Foundation of China
under the project No. 10505010 and MOE of China under project No.~IRT0624.
The work of T.H. was partly supported by Grant-in-Aid for Scientific
Research No.~19740130 and by Sumitomo Foundation No.~080734. 
\end{acknowledgments}

\end{document}